\documentclass[conference]{IEEEtran}
\usepackage{graphicx}
\usepackage{url}
\usepackage{stfloats}
\usepackage{amsfonts,amssymb,amsmath,bm,paralist,theorem,cite,ifthen,color}
\usepackage{array}
\usepackage{calc}
\usepackage[format=hang]{subfig}
\usepackage{cases}
\usepackage{algorithm}
\usepackage{algorithmic}

\newlength{\picwidth}
    \makeatletter
    \def\multilimits@{\bgroup
  \Let@
  \restore@math@cr
  \default@tag
 \baselineskip\fontdimen10 \scriptfont\tw@
 \advance\baselineskip\fontdimen12 \scriptfont\tw@
 \lineskip\thr@@\fontdimen8 \scriptfont\thr@@
 \lineskiplimit\lineskip
 \vbox\bgroup\ialign\bgroup\hfil$\m@th\scriptstyle{##}$\hfil\crcr}
    \def\Sb{_\multilimits@}
    \def\endSb{\crcr\egroup\egroup\egroup}
\makeatother

{\begin{list}{}%
    {\setlength{\rightmargin}{0pc}%
    \setlength{\leftmargin}{1pc} }} %
{\end{list}}

\newlength{\twidth}
\newtheorem{Prop}{Proposition} 

\theorembodyfont{\rmfamily}

\newtheorem{Fact}{Fact}

\ifCLASSINFOpdf
\else
\fi
    \IEEEoverridecommandlockouts

\begin{document}
%
\title{\huge A Decentralized Method for Joint Admission Control and Beamforming in Coordinated Multicell Downlink}

\author{\IEEEauthorblockN{\it Hoi-To Wai, Wing-Kin Ma} \\
\IEEEauthorblockA{Department of Electronic Engineering, \\
 The Chinese University of Hong Kong, Shatin, N.T., Hong Kong\\
Email: \url{htwai@ee.cuhk.edu.hk,wkma@ieee.org}}
\vspace{-0.7cm}~
\thanks{This work was supported by a Direct Grant by the Chinese University of Hong Kong (Project ID 2050506).}
 }

\maketitle

\begin{abstract}
%
In cellular networks, admission control and beamforming optimization are intertwined problems. While beamforming optimization aims at satisfying users' quality-of-service (QoS) requirements or improving the QoS levels, admission control looks at how a subset of users should be selected so that the beamforming optimization problem can yield a reasonable solution in terms of the QoS levels provided. However, in order to simplify the design, the two problems are usually seen as separate problems. This paper considers joint admission control and beamforming (JACoB) under a coordinated multicell MISO downlink scenario. We formulate JACoB as a user number maximization problem, where selected users are guaranteed to receive the QoS levels they requested. The formulated problem is combinatorial and hard, and we derive a convex approximation to the problem. A merit of our convex approximation formulation is that it can be easily decomposed for per-base-station decentralized optimization, namely, via block coordinate decent. The efficacy of the proposed decentralized method is demonstrated by simulation results.
\end{abstract}


\IEEEpeerreviewmaketitle

\vspace*{-0.5\baselineskip}
\section{Introduction} \vspace{-0.1cm}

Coordinated beamforming (CoBF) \cite{gesbert10,wyu10} is a recently studied technique to mitigate intercell interference (ICI) in the downlink of multicell cooperative systems.
In CoBF, the neighboring BSs share the same frequency band and employ beamforming for data transmission. The transmit beamformers at different BSs are coordinately designed according to the channel conditions and certain design formulation, e.g., maximum system throughput, minimum transmit power, to name a few.
Compared to fully multicell cooperative techniques such as network MIMO \cite{gesbert10},
CoBF has an advantage that the BS cooperation overheads are not as significant, and
yet appealing performance may be achieved.

Meanwhile, admission control also plays an important role in cellular systems.
As cellular systems are usually congested, with lots of users awaiting service, it is necessary for the BSs to decide which user is served or not.
Admission control refers to methods of selecting users.
While admission control and beamforming are commonly seen as two separate problems,
they are fundamentally dependent on each other.
Recent work has demonstrated that by considering admission control and beamforming jointly, promising system performance can be achieved \cite{luo08,Liu2012}.
However, joint admission control and beamforming (JACoB) is a challenging problem.
It has been shown that JACoB is NP-hard even under a single-cell scenario~\cite{luo08}.
Hence, as a compromise, one may consider approximation approaches.

This paper describes a JACoB approach to CoBF in multicell MISO downlink.
A distinguishing part of the present work is that our proposed JACoB formulation can be easily decomposed for decentralized implementation,
and the decentralized process is considered even more straightforward than those in CoBF (without admission control), e.g. \cite{wyu10,Tolli11}.
A key idea of our approach is to use the now popularized $\ell_1$ approximation method.
We very recently note that in a concurrent work \cite{Liu2012}, the authors have studied $\ell_1$ approximation for joint admission control and power control (i.e., no beamforming).
Nevertheless, the work \cite{Liu2012} does not investigate the multicell CoBF scenario and, more importantly, decentralized optimization considered here.
It is also worthwhile to mention \cite{luo08}, which considers JACoB under a single cell scenario.
The formulation used there is based on  a mixed-integer program formulation, and is processed by semidefinite relaxation (SDR).
While it is not difficult to see that the idea of \cite{luo08} can be extended to the multicell scenario,
one needs to assume centralized optimization and presently there is no reported work on how the method in \cite{luo08} can be decentralized.
In our simulations, we will show that our decentralized method yields a performance quite on a par with the centralized method in \cite{luo08}.

\vspace{-0.1cm}
\section{System model} \vspace{-0.1cm}

Consider a cellular system with $M$ coordinating BSs.
Each BS is equipped with $N$ transmit antennas.
In each cell, there are $K$ single-antenna user terminals;
thus the total number of users in the system is $KM$.
The set of users associated with the $i$th cell, or the $i$th BS,
is denoted by $\mathcal{K}_i \subset \mathcal{K} = \{1,2,...,KM\}$.
We assume that $\mathcal{K} = \mathcal{K}_1 \cup \cdots \cup \mathcal{K}_M$ and $\mathcal{K}_i \cap \mathcal{K}_j = \emptyset$ whenever $ i \neq j$,
i.e., each user is served only by one BS.
The scenario of interest is downlink, with an emphasis on CoBF.
Assuming that the BS-to-user channels are frequency-flat and slow,
and that the linear unicast transmit beamforming scheme is employed,
we can characterize the CoBF system performance by the received signal-to-interference-and-noise ratios (SINRs)
(for more complete system model descriptions, see the literature, such as \cite{wyu10}):
\begin{equation} \label{eq:sinr}
\hspace{-.00cm} {\sf SINR}_{q} \hspace{-.1cm} = \hspace{-.1cm} \displaystyle \frac{ | {\bm h}_{i(q),q}^H {\bm w}_{q} |^2 } {  \displaystyle \sigma_{q}^2 + \hspace{-.45cm} \sum_{m \in \mathcal{K}_{i(q)} \setminus \{q\} } \hspace{-.45cm} |{\bm h}_{i(q),q}^H {\bm w}_{m} |^2 +  \hspace{-.2cm} \sum_{j \neq i(q)} \sum_{m \in \mathcal{K}_j } \hspace{-.1cm} |{\bm h}_{j,q}^H {\bm w}_{m} |^2 }, \hspace{-.1cm}
\end{equation}
where
$q \in \{ 1, 2, \ldots, KM \}$ is the user index,
$i(q) \in \{1,...,M\}$ denotes the BS with which the $q$th user is associated
(i.e., $i(q)$ is such that $q \in \mathcal{K}_{i(q)}$),
${\bm h}_{j,q} \in \mathbb{C}^{N}$ is the channel response from the $j$th BS to the $q$th user,
$\sigma_{q}^2$ is the noise variance,
and ${\bm w}_{q} \in \mathbb{C}^N$ is the beamforming vector for the $q$th user.

The difference between the SINR model in \eqref{eq:sinr} and a single-cell-based SINR model is that the former explicitly models the ICI,
which is given by the third term in the denominator of \eqref{eq:sinr}.
On the contrary, in the single-cell case,
the ICI is usually treated as a constant, and is absorbed by $\sigma_q^2$.
The idea of CoBF is therefore to ask the BSs to coordinately design $\{ {\bm w}_{q} \}_{q=1}^{KM}$, so that ICI may be jointly mitigated.

To motivate the study of JACoB, let us quickly review a CoBF design problem, specifically, the design proposed in \cite{wyu10}.
In that design, the BSs jointly design the beamforming vectors $\{ {\bm w}_{q} \}_{q=1}^{KM}$
such that the SINR of each user is no less than a user-requested threshold $\gamma_q$.
Moreover, the design objective aims at minimizing the total transmit power.
This amounts to the following design optimization problem:
\begin{equation} \label{eq:mcbf}
\begin{array}{rl}
\displaystyle \min_{ \{ {\bm w}_{q} \}_{q=1}^{KM} } & \sum_{q=1}^{KM} \| {\bm w}_q \|_2^2 \\
{\rm s.t.} & {\sf SINR}_{q} \geq \gamma_{q},~q=1,2,...,KM, \\
& \sum_{q \in \mathcal{K}_i} \| {\bm w}_{q} \|_2^2 \leq P_{max,i},~i=1,...,M,
\end{array}
\end{equation}
where the last constraints in \eqref{eq:mcbf} are per-BS power budget constraints,
with $P_{max,i}$ specifying the maximum transmit power of the $i$th BS\footnote{As a minor point to note, the previous study \cite{wyu10} does not incorporate the per-BS power budget constraints.},
and $\| \cdot \|_2$ is the $\ell_2$ norm.
At first glance problem~\eqref{eq:mcbf} seems to be nonconvex, since the SINR constraints are nonconvex in $\{ {\bm w}_{q} \}_{q=1}^{KM}$.
Actually, problem~\eqref{eq:mcbf} can be solved in a convex and tractable fashion, using either the second-order cone programming formulation or the semidefinite relaxation (SDR) formulation; see \cite{wyu10} and \cite{mats10} for more detail.

\begin{figure}[t]
\begin{center}
\resizebox{0.95\linewidth}{!}{%
\includegraphics{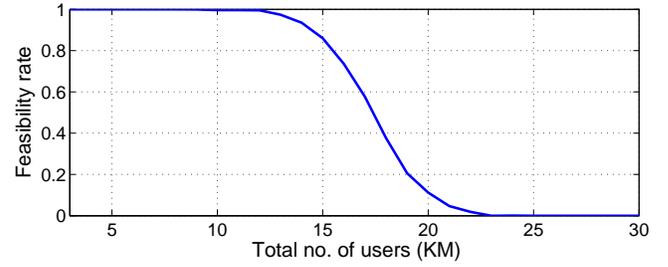}}
\end{center}
\caption{The feasibility rate of the CoBF problem \eqref{eq:mcbf}. $M=3$, $N=8$, $\gamma_q= 6$ dB, and $P_{max,i}$ = 46 dBm.} \label{fig:feasible}
\vspace{-0.3cm}
\end{figure}

Cellular systems are usually congested, with lots of users awaiting service.
A subsequent issue relevant to the CoBF problem~\eqref{eq:mcbf} is that we may be unable to find a beamforming solution $\{ {\bm w}_{q} \}_{q=1}^{KM}$ that satisfies all the users' SINR requests.
In other words, {\it problem~\eqref{eq:mcbf} may be infeasible}.
To illustrate this issue,
we simulated the feasibility rate of problem~\eqref{eq:mcbf} against the total number of users $KM$. The simulation result is plotted in Figure~\ref{fig:feasible}.
The feasibility rate was evaluated by counting the number of instances for which \eqref{eq:mcbf} is feasible, under randomly generated channels.
We observe that problem \eqref{eq:mcbf} has a low feasibility rate when the number of users is large.

\section{Joint admission control and beamforming}
This work considers joint admission control and beamforming (JACoB). The problem is stated as follows:
\vspace{.2cm}

\noindent \fbox{ \parbox{0.95\linewidth}{ \vspace{.05cm}
\textbf{Joint admission control and beamforming (JACoB)}:
\vspace{.1cm}

\emph{Select a maximum number of users,
such that there exists a beamforming solution $\{{\bm w}_q\}$
that satisfies all the selected users' SINR requests.}\vspace{.05cm}
}} \vspace{.2cm}

In the following, we will first provide an optimization formulation for JACoB,
and derive a convex approximation to the formulated problem.
Then, a decentralized method based on the convex approximation will be developed.

\subsection{Centralized method for JACoB}

Our endeavor starts with formulating JACoB in a mathematically convenient form.
First of all, let
\[ {\bm W}_q = {\bm w}_q {\bm w}_q^H, \quad q=1,\ldots,KM, \]
and observe the following equivalence (cf. \eqref{eq:sinr}):
\begin{equation}\label{eq:equiv} {\sf SINR}_q \geq \gamma_q \Longleftrightarrow 0 \geq f_q( \{ {\bm W}_{m} \}_{m=1}^{KM} ), \end{equation}
where we define
\[ \begin{array}{l} \displaystyle f_{q} ( \{ {\bm W}_{m} \}_{m=1}^{KM} ) \triangleq 1 + \sum_{j \neq i(q)} {\rm Tr}\Big( {\bm H}_{j,q} \Big( \sum_{m \in \mathcal{K}_j } {\bm W}_{m} \Big) \Big) \\
\displaystyle \hfill + {\rm Tr} \Big( {\bm H}_{i(q),q} \Big( \sum_{m \in \mathcal{K}_{i(q)} \setminus \{q \}} {\bm W}_{m} - \frac{1}{\gamma_{q}} {\bm W}_{q} \Big) \Big),
\end{array} \]
and ${\bm H}_{j,q} \triangleq {\bm h}_{j,q} {\bm h}_{j,q}^H / \sigma_q^2$.
We claim that JACoB can be formulated as the following $\ell_0$ minimization problem:
\begin{subequations}  \label{eq:mcadm}
\begin{align}
\displaystyle \min_{ {\bm t},\{ {\bm W}_q \}_{q=1}^{KM} } &~~ \| {\bm t} \|_0+  \textstyle \epsilon \sum_{q=1}^{KM} {\rm Tr}({\bm W}_q) \label{eq:mcadm_l1a} \\
{\rm s.t.} &~~ \textstyle \sum_{q \in \mathcal{K}_i} {\rm Tr}({\bm W}_{q}) \leq P_{max,i},~i=1,...,M, \label{eq:mcadm_l1b} \\
& ~~{\bm W}_{q} \succeq {\bm 0},~\forall~q, \label{eq:mcadm_l1c} \\
&~~ t_q = \max\{0,f_{q} ( \{ {\bm W}_{m} \}_{m=1}^{KM} )\},~\forall~q, \label{eq:mcadm_l1d} \\
&~~{\rm rank}({\bm W}_q) = 1, ~\hfill~\forall~q,  \label{eq:mcadm_l1e}
\end{align}
\end{subequations}
where $0 < \epsilon < 1 / \sum_{i=1}^{M} P_{max,i}$ is a penalty parameter,
$\bm{W}_q \succeq \bm{0}$ means that $\bm{W}_q$ is positive semidefinite,
and $\| \bm{t} \|_0$ is the $\ell_0$ norm, which counts the number of nonzero elements in $\bm{t}$.

Let us describe why problem \eqref{eq:mcadm} delivers the above defined JACoB goal.
Firstly, constraints \eqref{eq:mcadm_l1c} and \eqref{eq:mcadm_l1e} are equivalent to ${\bm W}_q = {\bm w}_q {\bm w}_q^H$.
Secondly, by substituting \eqref{eq:mcadm_l1d} into the first term of \eqref{eq:mcadm_l1a}, i.e., $\| {\bm t} \|_0$,
and observing \eqref{eq:equiv},
we can see that $\| {\bm t} \|_0$ is counting the number of unserved or unadmitted users.
Hence, if we ignore the second term of \eqref{eq:mcadm_l1a}, then
problem \eqref{eq:mcadm} minimizes the number of unadmitted users.
Thirdly, the second term of \eqref{eq:mcadm_l1a}, i.e, $\epsilon \sum_{q=1}^{KM} {\rm Tr}({\bm W}_q)$, is a penalty term.
It is used to encourage more power-efficient beamforming solutions.
It can be shown that problem \eqref{eq:mcadm} with $\epsilon= 0$ (i.e., direct unadmitted user minimization) achieves the same number of unadmitted users as problem \eqref{eq:mcadm} with $\epsilon <  1 / \sum_{i=1}^{M} P_{max,i}$.

Problem \eqref{eq:mcadm} is difficult to solve.
As a remedy, we adopt a convex approximation approach.
Our approximation involves two steps.
First, we replace the hard $\ell_0$ norm function by the $\ell_1$ norm,
which is now a popularized trick in compressive sensing.
Second, we remove the rank-one constraints \eqref{eq:mcadm_l1e},
which is well known as SDR~\cite{mats10}.
The above two approximations lead us to the following $\ell_1$ approximate JACoB problem:
\begin{subequations}  \label{eq:mcadm_l1}
\begin{align}
\displaystyle \hspace{-.4cm} \min_{ {\bm t},\{ {\bm W}_q \}_{q=1}^{KM} } & ~~\textstyle \| {\bm t} \|_1+  \epsilon \sum_{q=1}^{KM} {\rm Tr}({\bm W}_q) \label{eq:mcadm_newl1_a} \\
{\rm s.t.} & ~~\textstyle \sum_{q \in \mathcal{K}_i} {\rm Tr}({\bm W}_q) \leq P_{max,i},~i=1,...,M, \label{eq:mcadm_newl1_b} \\
& \textstyle ~~ t_q \geq 0,~t_q \geq f_{q} ( \{ {\bm W}_{m} \}_{m=1}^{KM} ), \label{eq:mcadm_newl1_c} \\
& ~~{\bm W}_q \succeq {\bm 0}, ~\forall~q, \label{eq:mcadm_newl1_d}
\end{align} \end{subequations}
where $\| \cdot \|_1$ is the $\ell_1$ norm.
Note that in \eqref{eq:mcadm_l1}, we replace \eqref{eq:mcadm_l1d}
by \eqref{eq:mcadm_newl1_c}, which can be easily verified to be equivalent.

The $\ell_1$ approximate JACoB problem \eqref{eq:mcadm_l1} is convex. In fact, problem \eqref{eq:mcadm_l1} can be written as an SDP.
Hence, for centralized implementation, we can solve problem \eqref{eq:mcadm_l1} by using a readily available SDP solver.
Moreover, we show that the second approximation, i.e., SDR, is a tight relaxation:
\begin{Prop}
For $\epsilon > 0$,
any optimal solution $\{ {\bm W}_q^\star \}_{q =1}^{KM}$ of problem~\eqref{eq:mcadm_l1}  must satisfy ${\rm rank}({\bm W}_q^\star) \leq1$ for all $q$.
\end{Prop}
The proof of Proposition~1 is skipped here owing to the limit of space.
The idea behind the proof is to examine the KKT conditions and exploit the rank-one structure of ${\bm H}_{i(q),q}$.
We should also note that Proposition~1 is different from the SDR tightness results in \cite{Palomar2010}, which may be seemingly similar at first look.
Simply speaking, \cite{Palomar2010} studies more general  ${\bm H}_{j,q}$ (which may take any rank), but may not solve the problem in Proposition~1.
Proposition~1 means that solving problem~\eqref{eq:mcadm_l1} automatically leads to a set of beamforming solutions (recall $\bm{W}_q = {\bm w}_q {\bm w}_q^H$ for rank-one positive semidefinite $\bm{W}_q$), and there is no loss in applying SDR.

\subsection{Decentralized method for JACoB}

A significant advantage of the $\ell_1$ approximate JACoB formulation in \eqref{eq:mcadm_l1} is that
it can be easily decomposed for decentralized optimization.
To see this, let
\[ \bm{\mathcal{W}}_i \triangleq \{ {\bm W}_m \}_{m \in \mathcal{K}_i},~i=1,...,M. \]
Notice that $\bm{\mathcal{W}}_i$ corresponds to the beamforming vectors controlled by the $i$th BS.
Now, by substituting \eqref{eq:mcadm_newl1_c} into \eqref{eq:mcadm_newl1_a},
we can reformulate \eqref{eq:mcadm_l1} as
\begin{equation} \label{eq:mcadm_bcd}
\begin{array}{rl}
\displaystyle \hspace{-.6cm} \min_{ \{ {\bm W}_q\}_{q=1}^{KM}} & \hspace{-.1cm}\displaystyle  \sum_{q=1}^{KM} \Big( \hspace{-.05cm} \max\{ 0,  f_{q} ( \bm{\mathcal{W}}_1,...,\bm{\mathcal{W}}_M ) \} + \epsilon {\rm Tr}( {\bm W}_q ) \Big) \vspace{-.0cm}  \\
{\rm s.t.} & \sum_{q \in \mathcal{K}_i} {\rm Tr}({\bm W}_q) \leq P_{max,i},~i=1,...,M,\\
& {\bm W}_q \succeq {\bm 0}, ~q=1,...,KM.
\end{array} \hspace{-.2cm}
\end{equation}
A unique feature with problem \eqref{eq:mcadm_bcd} is that the constraints are per-BS decoupled (note that this is not the case with the CoBF problem in \eqref{eq:mcbf}).
As a result, we can directly apply per-BS alternating optimization.
To be specific, we employ block coordinate descent (BCD).
In BCD, we update only one beamforming block $\bm{\mathcal{W}}_i$, while holding the other blocks fixed.
This BCD update is done cyclically with respect to the BSs, until some stopping rule is satisfied.

A curious question is whether the above-described BCD method would converge to the optimum of problem \eqref{eq:mcadm_bcd}.
Unfortunately, this may not be guaranteed---BCD may not converge to the optimum for problems whose objective functions are {\it not} continuously differentiable, even if the problem is convex~\cite{bertsekas99}.
The function $\max\{0,x\}$ seen in problem \eqref{eq:mcadm_bcd} exactly falls into this case.
To remedy this, we apply a smooth approximation to \eqref{eq:mcadm_bcd} using the one-sided Huber function
\[
h(x) = \begin{cases}
0 & ,~{\sf if}~x \leq 0, \\
\displaystyle 0.5 x^2 & ,~{\sf if}~0 <  x \leq 1,\\
\displaystyle x - 0.5 & ,~{\sf if}~x > 1.
\end{cases}
\]
The Huber function $h(x)$ is continuously differentiable in $x$. Applying the approximation $\max\{0,x\} \approx h(x)$,
we obtain the Huber approximate JACoB problem:
\begin{equation} \label{eq:mcadm_huber}
\begin{array}{rl}
\displaystyle \hspace{-.3cm} \min_{ \{ {\bm W}_q\}_{q=1}^{KM}} & \hspace{-.2cm}  \displaystyle \sum_{q=1}^{KM} \Big( h(  f_{q} ( \bm{\mathcal{W}}_1,...,\bm{\mathcal{W}}_M ) ) + \epsilon {\rm Tr}( {\bm W}_q ) \Big) \\
{\rm s.t.} & \sum_{q \in \mathcal{K}_i} {\rm Tr}({\bm W}_q) \leq P_{max,i},~i=1,...,M,\\
& {\bm W}_q \succeq {\bm 0}, ~q=1,...,KM.
\end{array}
\end{equation}
From this point on, we will concentrate on the BCD of problem~\eqref{eq:mcadm_huber}.

Let us consider the BCD update of problem~\eqref{eq:mcadm_huber} with respect to the $i$th block $\bm{\mathcal{W}}_i$,
holding the other blocks $\{ \hat{\bm{\mathcal{W}}}_j \}_{j \neq i} \triangleq \{ \hat{\bm W}_m \}_{m \notin \mathcal{K}_i}$ fixed.
The respective problem is
\[
\begin{array}{rl}
\displaystyle \hspace{-.3cm} \min_{ \{ {\bm W}_m\}_{m \in \mathcal{K}_i}} & \hspace{-.1cm}   \displaystyle \sum_{q=1}^{KM} h( f_{q} ( \bm{\mathcal{W}}_i , \{ \hat{\bm{\mathcal{W}}_j} \}_{j \neq i} ) ) + \hspace{-.1cm} \sum_{m \in \mathcal{K}_i} \hspace{-.1cm}  \epsilon {\rm Tr}( {\bm W}_m ) \vspace{.2cm} \\
{\rm s.t.} & \sum_{m \in \mathcal{K}_i} {\rm Tr}({\bm W}_m) \leq P_{max,i},~{\bm W}_m \succeq {\bm 0},~m\in \mathcal{K}_i, \\
\end{array}
\]
which can be expressed as a convex problem (see, e.g., \cite{boyd04}):
\begin{subequations}  \label{eq:mcadm_bcd_1}
\begin{align}
\displaystyle \hspace{-.3cm} \min & ~~ \displaystyle \sum_{q=1}^{KM} \left(\frac{1}{2}u_q^2 + v_q \right) + \hspace{-.1cm} \sum_{m \in \mathcal{K}_i} \hspace{-.1cm}  \epsilon {\rm Tr}( {\bm W}_m ) \vspace{.2cm} \\
{\rm s.t.} & ~~u_q + v_q \geq f_q( \bm{\mathcal{W}}_i, \{ \hat{\bm{\mathcal{W}}}_j \}_{j \neq i} ),~\forall~q, \label{eq:mcadm_bcd_b} \\
& ~~u_q,~v_q \geq 0,~\forall~q, \\
& \sum_{m \in \mathcal{K}_i} {\rm Tr}({\bm W}_m) \leq P_{max,i},~{\bm W}_m \succeq {\bm 0},~m \in \mathcal{K}_i.
\end{align}
\end{subequations}
Constraints \eqref{eq:mcadm_bcd_b} seems to indicate that full knowledge of $\{ \hat{\bm{\mathcal{W}}}_j \}_{j \neq i}$ is required, in order to solve the BCD update \eqref{eq:mcadm_bcd_1}.
Actually, this may be not necessary.
Notice that for $q \in \mathcal{K}_i$, constraint \eqref{eq:mcadm_bcd_b} can be expressed as:
\[
u_q + v_q \geq 1 + \sum_{j \neq i} \hat{\Omega}_{j,q} + {\rm Tr} \Big( {\bm H}_{i,q} \Big( \sum_{ m \in \mathcal{K}_i \setminus \{q\} } {\bm W}_m - \frac{ {\bm W}_q }{\gamma_q} \Big) \Big),
\]
and for $q \notin \mathcal{K}_i$,
\[
u_q + v_q \geq 1 + \sum_{j \neq i} \hat{\Omega}_{j,q} + {\rm Tr} \Big( {\bm H}_{i,q} \Big( \sum_{ m \in \mathcal{K}_i } {\bm W}_m \Big) \Big),
\]
where $\hat{\Omega}_{j,q}$ are scalar constants defined as:
\[
\hat{\Omega}_{j,q} = \begin{cases}
\displaystyle {\rm Tr} \Big( {\bm H}_{j,q} \Big( \sum_{m \in \mathcal{K}_j \setminus \{q \}} \hat{\bm W}_{m} - \frac{\hat{\bm W}_{q}}{\gamma_{q}}  \Big) \Big) &,~q \in \mathcal{K}_j, \\
\displaystyle {\rm Tr} \Big( {\bm H}_{j,q} \Big( \sum_{m \in \mathcal{K}_j} \hat{\bm W}_{m}  \Big) \Big) &,~q \notin \mathcal{K}_j.
\end{cases}
\]
Hence, if the $i$th BS knows i) the matrices $\{ {\bm H}_{i,q} \}_{q=1}^{KM}$, i.e., the channel response from the $i$th BS to the users in the system; and ii) the scalar constants $\{ \hat{\Omega}_{j,q} \}_{j\neq i, q \in \mathcal{K}}$, then problem \eqref{eq:mcadm_bcd_1} can be solved independently at the $i$th BS.
In fact, the first premise can be satisfied automatically as it can be assumed that each BS knows the channel response from itself to the users in the system \cite{wyu10}.
To satisfy the second premise, we can utilize the backhaul link between the BSs. Specifically, the scalars $\{ \hat{\Omega}_{i,q} \}_{q \in \mathcal{K}}$ can be computed and broadcast to the other BSs after the $i$th BCD update is solved.
There are $KM$ real numbers to be broadcast at each iteration.
This justifies our claim that the $i$th BS can solve \eqref{eq:mcadm_bcd_1} alone. The BCD method for \eqref{eq:mcadm_huber} is summarized in Algorithm~\ref{alg:bcd}.

Recall that the reason for employing the Huber function in \eqref{eq:mcadm_huber} is to provide a smooth approximation to JACoB,
avoiding the original nondifferentiable objective function which may result in BCD non-convergence problems.
But can the smooth approximation guarantee convergence to the optimum?
By invoking an available BCD convergence analysis result~\cite{Grippo2000}, we have the following claim:
\begin{Fact} \cite{Grippo2000}
{\it The sequence $\{ \{{\bm W}_q^{(k)}\}_{q=1}^{KM} \}_k$ generated by Algorithm~\ref{alg:bcd} has limit points and every limit point of the sequence $\{ \{{\bm W}_q^{(k)}\}_{q=1}^{KM} \}_k$ is an optimal solution to \eqref{eq:mcadm_huber}.} \vspace{0.2cm}
\end{Fact}
Readers are referred to Proposition~6 in \cite{Grippo2000} for more detail.
We should note that the important premises for us to use this available result are that the objective function of \eqref{eq:mcadm_huber} is convex, continuously differentiable in $\{ \bm{\mathcal{W}}_i \}_{i=1}^M$ and the constraint set for each $\bm{\mathcal{W}}_i$ is convex and compact.
Furthermore, by extending Proposition~1, we can prove that the intermediate solutions $\{ {\bm W}_q^{(k)} \}_{q \in \mathcal{K}_i}$ in Algorithm~1 at the $i$th BCD update are always of rank-one:
\begin{Prop}
For $\epsilon > 0$ and for each $i$, any optimal solution $\{ {\bm W}_q^{(k)} \}_{q \in \mathcal{K}_i}$ of the $i$th BCD update in Algorithm~1 must satisfy ${\rm rank}({\bm W}_q^{(k)}) \leq1$ for all $q$ in $\mathcal{K}_i$.
\end{Prop}

{\algsetup{indent=1em}
\begin{algorithm}[t!]
  \caption{Block coordinate descent method for \eqref{eq:mcadm_huber}}\label{alg:bcd}
  \begin{algorithmic}[1]
    \REQUIRE initialization - $\{ {\bm W}_q^{(0)} \}_{q \in \mathcal{K}}$.
\STATE $k=1$;
\STATE For each $i=1,2,...,M$, the $i$th BS computes $\{ \hat{\Omega}_{i,q} \}_{q \in \mathcal{K}}$ and broadcasts them to the other BSs;
    \REPEAT
\FOR{$i=1$ to $M$}
\STATE The $i$th BS solves \eqref{eq:mcadm_bcd_1} given $\{ \hat{\Omega}_{j,q} \}_{j\neq i, q \in \mathcal{K}}$ to obtain $\{ {\bm W}_q^{(k)} \}_{q \in \mathcal{K}_i}$. The scalars $\{ \hat{\Omega}_{i,q} \}_{q \in \mathcal{K}}$ are computed and broadcast to the other BSs;
\ENDFOR
\STATE $k= k +1$;
\UNTIL convergence.
\RETURN an optimal solution to \eqref{eq:mcadm_huber} - $\{ {\bm W}_q^{(k)} \}_{q \in \mathcal{K}}$.
  \end{algorithmic}
\end{algorithm}}

\section{Deflation heuristic}
Both $\ell_1$ and Huber approximate JACoB problems (cf. problems \eqref{eq:mcadm_l1} and \eqref{eq:mcadm_huber}, respectively) can be seen as some kind of ``soft decision'' formulations for handling admission control. In order to select more users for service, we can apply a hard decision using the deflation heuristic.
Similar to \cite{luo08,Liu2012}, the heuristic is initialized by considering all users in the system, then the users are dropped one-by-one.
At first the BSs solve \eqref{eq:mcadm_l1} or \eqref{eq:mcadm_huber} either centrally or using the BCD method.
Our user dropping rule is based on the value of $t_q^\star \triangleq \max\{0, f_q(\cdot)\}$ which relates directly to the satisfiability of the SINR threshold for user $q$. The user with the largest $t_q^\star$ will be dropped.

When the number of users in the system is too large, we may encounter cases where the optimal solution to \eqref{eq:mcadm_l1} or \eqref{eq:mcadm_huber} is trivial, i.e., ${\bm W}_q = {\bm 0}$ for all $q$.
Here we state an easy-to-check condition for identifying such cases.

\vspace{-.3cm}
\begin{Fact}
(Prescreening condition)\footnote{A similar condition has been discovered recently in \cite{Liu2012} for the joint power and admission control problem. Our results applies to the case with CoBF.}
If
\begin{equation}\label{eq:prescreen}
\bm{\Phi}_q ( \{{\bm H}_{i(q),m}\}_{m \in \mathcal{K}} ) )  \succeq {\bm 0},~\forall~q \in \mathcal{K},
\end{equation}
then solving \eqref{eq:mcadm_l1} or \eqref{eq:mcadm_huber} gives a trivial solution, i.e., ${\bm W}_q = {\bm 0}$ for all $q$ in $\mathcal{K}$, where
\[
\bm{\Phi}_q ( \{{\bm H}_{i(q),m}\}_{m \in \mathcal{K}} ) ) \triangleq \epsilon {\bm I} +  \sum_{m \neq q} {\bm H}_{i(q),m} - \frac{1}{\gamma_q} {\bm H}_{i(q),q}.
\]
\end{Fact} \vspace{-.2cm}
The proof is omitted due to space limitation.
Inspired by fact~2, we now adopt a \emph{prescreening procedure} where we drop the users gradually until condition \eqref{eq:prescreen} gets violated. Specifically, at each time, we remove user $q$ with $\bm{\Phi}_q(\cdot )$ that gives the largest minimum eigenvalue
As condition \eqref{eq:prescreen} can be checked in closed-form, the prescreening procedure can be run at a low complexity.
The deflation heuristic, together with prescreening, are summarized as follows\footnote{Note that both the deflation heuristic and prescreening procedure can be operated in a decentralized manner.}:

\vspace{.2cm}

\noindent \fbox{
\parbox{0.95\linewidth}{\underline{\textbf{Deflation heuristic:}}

\begin{enumerate}
\item[$\bullet$] {\bf Initialize}: a set of users requesting service - $\mathcal{K} = \{1,2,...,KM\}$.
\item (Prescreening) Check condition \eqref{eq:prescreen}. If it holds, then remove user $m$ from $\mathcal{K}$ according to
$ m  = \arg \max_q \lambda_{min} ( \bm{\Phi}_q ( \cdot ) ) $
and repeat 1). Otherwise, go to 2).
\item (Deflation) Solve \eqref{eq:mcadm_l1} or \eqref{eq:mcadm_huber} for $\{ t_q^\star, {\bm W}_q^\star \}_{q \in \mathcal{K}}$. If $t_q^\star = 0$ for all $q \in \mathcal{K}$, terminate. Otherwise remove user $m$ from $\mathcal{K}$ according to $m = \arg \max_q t_q^\star$, repeat 2).
\item[$\bullet$] {\bf Return}: a set of selected users $\mathcal{K}$ and rank-one matrices $\{ {\bm W}_q^\star \}_{q \in \mathcal{K}}$ that decomposes into  beamforming vectors satisfying the SINR requirements.
\end{enumerate}
}
}
\vspace{.1cm}

\section{Numerical results}
This section presents numerical results for the proposed JACoB methods. The simulation environment is similar to \cite{wyu10}. We focus on a multicell scenario with 3 coordinating BSs where each BS is separated from the others by 2.8 km. For each simulation trial, the users' positions and their respective channels are randomly generated. The users are separated from their respective BS by at least 0.7 km and are assigned to the nearest BS. The channel is assumed to experience both small-scale and large-scale fading. The noise variance $\sigma_q^2$ is -92 dBm and the receive antenna gain is 5 dBi.
There are 8 transmit antennas and 15 users are assigned to each BS (i.e., 45 users awaiting service in total).

Here, the centralized method and the decentralized method refer to the deflation heuristic which uses a centralized solver for \eqref{eq:mcadm_l1} and the decentralized BCD method for \eqref{eq:mcadm_huber}, respectively. The scalar $\epsilon$ is chosen as $10^{-5}$ and the BCD method terminates when the relative change in objective value is less than $10^{-2}$.

\begin{figure}[t]
\begin{center}
\resizebox{0.95\linewidth}{!}{%
\includegraphics{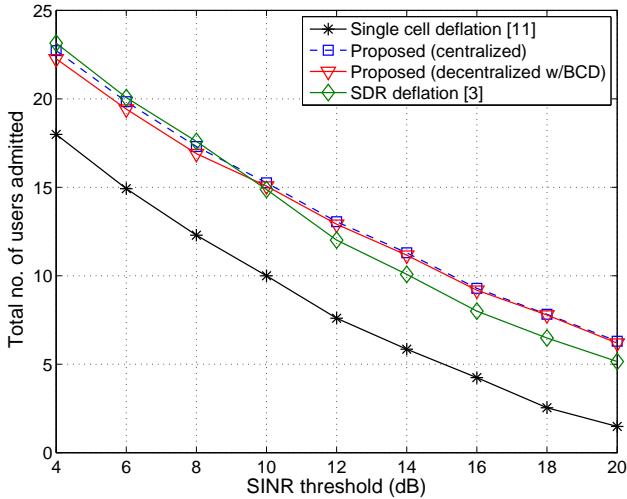}} \vspace{-.2cm}
\end{center}
\caption{Total no.~of users admitted, $M=3$, $N=8$, $K=15$.} \label{fig:user}
\vspace{-1.4cm}
\end{figure}

Figure~\ref{fig:user} shows the performance of different JACoB methods in terms of the total number of users selected for service. Here, two benchmarking methods are compared.
The ``SDR deflation method'' is adopted from \cite{luo08} and modified to operate in the multicell scenario. Note that it is a centralized method which may not be decomposed straightforwardly.
The ``single cell deflation method''
is a modified algorithm from \cite{Huh2010}, which is originally proposed as a suboptimal alternative to the CoBF problem \eqref{eq:mcbf}. The main feature of \cite{Huh2010} is that the ICI levels are always constrained below a fixed threshold, and therefore the beamforming design can be done independently at each BS.
The subsequent single-cell beamforming design is processed by our JACoB method.
Note that the resultant method can be implemented in a per-BS decentralized manner \emph{without} any BS coordination.

Turning back to Figure~\ref{fig:user}, we compare the performance of \cite{luo08} with the proposed centralized method. We observe that there are some performance gains with the proposed method in the high SINR regime.
Furthermore, the proposed decentralized method achieves a performance on a par with its centralized counterpart. The decentralized method should also be compared to the single cell deflation method, where the numerical results has clearly demonstrated the benefits of allowing BSs coordination.

Table~\ref{tab:iter} demonstrates the efficacies of the proposed decentralized BCD method with prescreening procedure in terms of the total number of iterations. The total number of iterations is defined as the {total number of BCD iterations} consumed \emph{throughout} the deflation heuristic, where multiple instances of \eqref{eq:mcadm_huber} are solved.
Note that the load on the backhaul link is directly proportional to the iteration count.
The iteration counts reported in Table~\ref{tab:iter} confirms that significant reduction in the number of iterations can be achieved \emph{with} the prescreening procedure.

\vspace{-.2cm}
\section{Conclusion}
The contributions of this paper are twofold.
First, we have developed a formulation of joint admission control and beamforming (JACoB) for coordinated multicell downlink,
wherein an efficient convex approach is proposed.
Second, we have built a decentralized JACoB method via a simple BCD procedure.
Simulation results have shown that the decentralized method achieves a performance on a par with the centralized method with fast convergence.

\begin{table}[!t]
\renewcommand{\arraystretch}{1.5}
\centering
\caption{Total no. of iterations, $M=3,N=8, K=15$.}
\label{tab:iter}
\begin{tabular}{|l||c|c|}
\hline
\bfseries Threshold ($\gamma$) & \bfseries With prescreening & \bfseries Without prescreening  \\
\hline \hline
12 dB & \bfseries 53.080 & 61.790 \\
\hline
20 dB & \bfseries 31.030 & 59.290 \\
\hline
\end{tabular} \vspace{-.3cm}
\end{table}

\bibliographystyle{IEEEtran}

\footnotesize
\bibliography{paper}

\end{document}